\documentstyle{article}

\begin{document}

\setlength{\parskip}{2ex}
\setlength{\parindent}{0em}
\setlength{\baselineskip}{4ex}

\newcommand{\be}{\begin{equation}}
\newcommand{\ee}{\end{equation}}
\newcommand{\bea}{\begin{eqnarray}}
\newcommand{\eea}{\end{eqnarray}}
\newcommand{\nn}{\nonumber}
\newcommand{\bean}{\begin{eqnarray*}}
\newcommand{\eean}{\end{eqnarray*}}
\newcommand{\half}{\frac{1}{2}}
\newcommand{\integ}[2]{\int\limits_{#1}^{#2}\!\!}
\newcommand{\del}{\partial}
\newcommand{\dslash}{\partial\hspace{-.55em} /}
\newcommand{\dm}{\partial_{\mu}}
\newcommand{\dn}{\partial_{\nu}}
\newcommand{\ds}{\partial_{\sigma}}
\newcommand{\dr}{\partial_{\rho}}
\newcommand{\gm}{\gamma_{\mu}}
\newcommand{\gn}{\gamma_{\nu}}
\newcommand{\gs}{\gamma_{\sigma}}
\newcommand{\gr}{\gamma_{\rho}}
\newcommand{\gum}{\gamma^{\mu}}
\newcommand{\gun}{\gamma^{\nu}}
\newcommand{\gus}{\gamma^{\sigma}}
\newcommand{\gur}{\gamma^{\rho}}
\newcommand{\e}{\varepsilon}
\newcommand{\gmn}{g^{\mu\nu}}
\newcommand{\gmnd}{g_{\mu\nu}}
\newcommand{\expect}[1]{\langle #1 \rangle}
\newcommand{\dd}{\mbox{d}}
\newcommand{\Gam}[1]{\Gamma\left(#1\right)}
\newcommand{\gE}{\gamma_E}
\newcommand{\ordo}[1]{{\cal O}(#1)}
\newcommand{\myref}[1]{(\ref{#1})}
\newcommand{\idk}{\int\frac{\dd^dk}{(2\pi)^d}}
\newcommand{\dhalf}{\frac{d}{2}}
\newcommand{\fourpi}{(4\pi)^{\frac{d}{2}}}
\newcommand{\la}{\lambda^a}
\newcommand{\lb}{\lambda^b}
\newcommand{\lc}{\lambda^c}
\newcommand{\ld}{\lambda^d}
\newcommand{\fabc}{f^{abc}}
\newcommand{\dabc}{d^{abc}}
\newcommand{\unit}{1}
\newcommand{\secI}[1]{\section{#1}}
\newcommand{\secII}[1]{\subsection{#1}}
\newcommand{\secIII}[1]{\subsubsection{#1}}
\newcommand{\jma}{j_{\mu}^a}
\newcommand{\jmb}{j_{\mu}^b}
\newcommand{\jmc}{j_{\mu}^c}
\newcommand{\jna}{j_{\nu}^a}
\newcommand{\jnb}{j_{\nu}^b}
\newcommand{\jnc}{j_{\nu}^c}
\newcommand{\jra}{j_{\rho}^a}
\newcommand{\CC}{C_{\mu\nu}^{\rho}}
\newcommand{\DD}{D_{\mu\nu}^{\sigma\rho}}
\newcommand{\bPsi}{\bar{\psi}}
\newcommand{\GN}{Gross-Neveu\ }

\title{
\begin{flushright} \small
hep-th/9708132  \\
ITP Budapest Report No. 534. 
\end{flushright} \vspace{1.5cm}
Systematic proof of the existence of Yangian symmetry in chiral Gross--Neveu
models }
\author{
  \normalsize{ Tamas Hauer\footnote{E-mail address: hauer@mit.edu}
    \footnote{Address: Center for Theoretical Physics, MIT, Cambridge, MA 02139 } }
  \\ \\
  \normalsize{ Institute for Theoretical Physics}\\
  \normalsize{ Lorand E\"otv\"os   University} \\
  \normalsize{ H-1088 Budapest, Puskin u. 5-7, Hungary}
}
\def\today{}
\maketitle

\begin{abstract}
The existence of non-local charges, generating a Yangian symmetry is
discussed in generalized chiral Gross-Neveu models. Their conservation can
be proven by a finite-loop perturbative computation, the order of which
is determined from group theoretic constants and is independent of the number
of flavors. Examples, where the 1-loop calculation is sufficient, include the
$SO(n)$-models and other more exotic groups and representations.  
\end{abstract}

 


\newpage


{\bf Introduction.} In the theory of two dimensional integrable systems
a fundamental role is 
played by those conserved quantities which guarantee the solvability of the
models. While in many cases classical integrability is manifest, the
proof of conservation and even the proper definition of these charges may be
quite subtle in the corresponding quantum theory. In certain models where the
symmetry algebra is generated by nonlocal charges
\cite{devega1,devega2} the approach 
originally proposed by L\"uscher \cite{Luscher} proved to be
fruitful. In these (asymptotically free) field theories, the
properties of nonlocal expressions of the local currents can be traced
back to the short distance singularities of the current algebra. The
existence of L\"uscher's nonlocal conserved charge is the consequence
of the fact that the operator product expansion (OPE) of the currents
close on themselves and their derivatives, which replaces, in the
quantum theory, the zero-curvature condition of the classical
currents. (See \cite{ope1} for a summary and \cite{AbAb} for a
complete review). The question about the closure of the current
algebra is a delicate one, and the answer varies from one model to
another. In the simplest case (like the O(n) non-linear sigma model
\cite{Luscher}, a large class of generalized sigma models
\cite{gensig} or the chiral SU(n) Gross-Neveu (GN)-model 
\cite{AbAb}) there are too few degrees of freedom to form 
operators which may ruin the conservation, while in other theories, extra
fields may be constructed and it is the dynamics of the model which determines
their ultimate presence or absence. For example, in the $CP^{N-1}$-model, 
the conservation is ruined by the extra term \cite{cpn1}, while
in its supersymmetric partner this quantum anomaly disappears
\cite{cpn2}. In a previous paper \cite{ope1} we studied the  $SU(n)$
(multiflavor), chiral Gross-Neveu model ((M)CGN)\cite{GN}, where -
thanks to renormalization group invariance - a one-loop perturbative
computation proved to be decisive and saved the desired form of the
current algebra. The aim of the present letter is to extend this
argument to a general class of chiral GN-type  models, which are defined
as Lagrangian field theories with current-current interaction, and where
the symmetry algebra generated by the currents is an {\em arbitrary
simple Lie-algebra}.  This family of theories (without
flavor multiplicity) was studied in
\cite{devega3} where, the non-local charges generating the classical
Yangian were constructed. Our goal is to investigate whether the
conservation of the non-local generator of the algebra survives the
quantization using the above strategy.  We will explicitly 
calculate the leading exponent of the perturbative coupling constant
of the OPE-coefficients in terms of group-theoretic constants and
obtain a simple formula (eq. \myref{alpha}), 
expressing the order of the needed perturbative calculation in terms
of quadratic Casimirs of certain representations. We also give a large
class of examples where zero- or one-loop results yield conclusion and
show that the multiplicity (flavor) of the fermion field does not
affect the question. These theories include the $SO(n)$
models\footnote{For the (one-flavor) $O(n)$-symmetric GN-models, non-local
Ward identities were proven in leading order of the $1/n$-expansion in  
\cite{zachos}} and a bunch of previously 
uninvestigated\footnote{This refers to the quantum case, the
study of the classical conservation laws is similar for the whole
family \cite{devega3}} models characterized by different groups,
representations and flavor multiplicity. 

The plan of the paper is as follows. First, we summarize the important
points in the connection between the non-local conservation laws and the OPE
of the current, then shortly review the RG-argument developed in \cite{ope1} 
showing how the anomalous dimension of certain operators play
fundamental role in the analysis. When generalizing to arbitrary groups, we
then express these quantities in terms of 
quadratic Casimirs of certain representations, deriving the formula
\myref{alpha}, which
determines the order of the needed perturbative calculation. Finally, we
present examples where no more then a 1-loop computation is sufficient to
reach conclusion.


{\bf OPE and Non-local Charge} In the field theories under consideration
we have a set of conserved, local currents: $\dm j^{a\mu}(x) = 0$,
transforming under the adjoint representation of 
a simple Lie-group, {\cal G}, with charges satisfying
\footnote{  $Q^a \equiv  \int dx j^{a0}(x,t) ; \;\;\; f^{abc}f^{bcd} =
-C_{adj}\delta^{ad}.$}:  
\bea 
\lbrack Q^a,Q^b \rbrack &=& \fabc Q^c. \label{algebra}
\eea
In addition we require the QFT be renormalizable and asymptotically free and
that all operators in the adjoint representation of {\cal G} have
higher canonical 
dimension than the current. If these conditions hold then, the general
form of the current-current OPE (up to vanishing terms as
$x\rightarrow 0$) is the following: 
\bea
\fabc \jmb(x)\jnc(0) &=& C^{\rho}_{\mu\nu}(x)\jra(0) +
                 D^{\sigma\rho}_{\mu\nu}(x)\ds\jra(0) + 
                 \sum_{i}E_i O_{i[\mu\nu]}^{a}(0),
\label{general} 
\eea 
where $C_{\mu\nu}^{\rho}(x)$ and $D_{\mu\nu}^{\sigma\rho}(x)$ are functions
with leading singularities $\ordo{|x|^{-1-0}}$ and $\ordo{|x|^{-0}}$ ($-0$ 
stands for logarithmic-like singularities) \cite{Luscher} which are
explicitly given in terms of one model-dependent scalar function,
$\xi(x^2)$ \cite{ope1}, while $E_i$ are {\em constant}
OPE-coefficients multiplying antisymmetric tensor operators, 
$O_{i[\mu\nu]}^{a}(x)$ \cite{cpn1,AbAb,ope1}. Using the OPE
\myref{general} one can prove \cite{Luscher,Lerr,cpn1} that, the
quantum analog of the classical charge
\bea
\label{classcharge}
Q_1^a = \frac{1}{4}\integ{-\infty}{\infty}
dy_1dy_2\epsilon(y_1-y_2)\fabc j_0^b(t,y_1)j_0^c(t,y_2) +
\integ{-\infty}{\infty}dy j_1^a(t,y)\, , 
\eea 
can always be consistently defined, but it is conserved only if all
the $E_i$'s are zero (both statements are independent of the concrete form of
$\xi(x^2)$).  Phrasing the condition in the above
form immediately yields the straightforward strategy for proving the
existence of the quantum charge: one assumes the presence of every
antisymmetric tensor operator allowed by the symmetries, and computes
the corresponding OPE-coefficients in some way; if all of them are zero
then the conservation is proved. This program was successfully carried
out in the $CP^{N-1}$-model with fermions, where supersymmetry
prevented the classical value from receiving quantum corrections
\cite{cpn2}; in our case renormalization group invariance will be of
great help.   


{\bf OPE and Renormalization Group}
As we showed in \cite{ope1}, the calculation of the OPE-coefficients
can be done using perturbation theory in models without
dimensionful parameters, since in this case renormalization group invariance
highly restricts the possible form of $E_i$. In the $SU(n)$ MCGN we faced one
extra operator and we expressed the leading exponent of its perturbative
expansion in terms of its 1-loop anomalous dimension, however -- as we will
see -- the argument is neither specific to $SU(n)$ nor is it
restricted to the extra operator standing alone. 

For simplicity, assume that the extra operators under consideration
renormalize multiplicatively (do not mix with others) and the anomalous 
dimension of $O_{i[\mu\nu]}^{a}$ is given by
\bea
{\cal D}\log(Z_i) &=&
\eta_{i,1} g^2 + \eta_{i,2} g^4 + \ldots,
\eea
where the lhs. is the renormalization-group derivative of the renormalization
constant $Z_i$, corresponding to $O_{i[\mu\nu]}^{a}$ and $g$ is the
perturbative coupling of the model. For the 
OPE-coefficient, 
\bea
E_i &=& E_{i,0}g^{2\alpha_i}(1+E_{i,1}g^2+E_{i,2}g^4+ \ldots),
\eea
the renormalization group equation yields the following relation between
$\alpha_i$, $\eta_{i,1}$, and the one-loop beta function coefficient,
$\beta_0$: 
\bea
\alpha_i &=& -\frac{\eta_{i,1}}{2\beta_0}.
\eea
This equation is the key point in the argument since $\alpha_i$ is either a
positive 
integer, in which case it determines the order of the perturbative calculation
needed to decide whether $E_i$ vanishes or not, or if it is negative or
non-integer then it does not allow $O_{i[\mu\nu]}^{a}$ to be present in the
OPE. 


{\bf Chiral Gross-Neveu models.} Now we turn to the CGN models, they are
defined by the following Minkowskian 
action: 
\bea
S = \int d^2x \left(\bPsi i\dslash\psi - \frac{g^2}{2}j_\mu^aj^{a\mu}\right).
\eea
The fermionic field, $\psi$ transforms under the irreducible representation,
$R$ (whe\-reas $\bPsi$ transforms under $\bar{R}$) of the simple (color)
Lie-group, {\cal G} and in case of the 
multiflavor  models it also carries a multiplicity (flavor) index. The current
in the interaction Lagrangian is defined as 
\bea
\jma(x)       &\equiv& \bar\psi(x)T^a\gamma_\mu\psi(x) ,
\eea
where $T^a$ are generator matrices in representation R (in
case of the multiflavor models summation over the multiplicity indices is
understood). In order not to get in conflict with our assumption that the
color current is the operator in the adjoint representation of {\cal G} of
the lowest canonical dimension, we require that the decomposition of
$\bar{R}\otimes R$ into irreducible representations does not contain the
adjoint more than once. 

The first steps in determining the operator content of the OPE
\myref{general} are: collecting the set of operators allowed by  
symmetry and canonical dimensional analysis; and then -- following our
method -- calculating their 1-loop renormalization. 
Apart from the current, there is one bilinear operator
in the adjoint representation, $i(\dm\bar\psi T^a\gamma_\nu\psi -
\bar\psi T^a\gamma_\nu\dm\psi)$ which, however, has opposite C-parity to
the current's and is excluded. Therefore canonical dimensional
analysis allows only for operators which are quadrilinear expressions of the
fermionic field. In order to identify these fields we search for operators in
the adjoint representation composed of the direct product 
$\bar{R}\otimes R\otimes\bar{R}\otimes R$ that is, we have to decompose this
quadratic product into irreducibles. The following sequence will prove to be
useful: decompose first $\bar{R}\otimes R$ (form bilinears) and then the
pairwise products (form bilinear of bilinears)
from the two direct sums and look for the adjoint representations:
\bea
(\bar{R}\otimes R)\otimes(\bar{R}\otimes R)=
(\ldots\oplus R_1\oplus\ldots)\otimes(\ldots\oplus R_2\oplus\ldots)=
\ldots\oplus R_{adj}\oplus\ldots 
\eea
It turns out that this decomposition will guarantee the multiplicative
renormalization and it is the Casimir of $R_1$ and $R_2$ ``defined'' above
that enters the anomalous dimension. To write down explicitly the operator we
have in mind, denote the projectors on the basis elements of $R_1$ and $R_2$
with $C^{1\alpha}$ and $C^{2\rho}$, respectively and the one on the basis
elements in the
adjoint representation by $h^a_{\alpha\rho}$:
\bea
O_{[\mu\nu]}^a &\equiv& h^a_{\alpha\rho}
       (\bPsi\gamma_{[\mu}C^{1\alpha}\psi)(\bPsi\gamma_{\nu]}C^{2\rho}\psi).
\eea
To calculate the anomalous dimension of this operator one has to compute the
one-loop four-particle correlation function. This contains four divergent
Feynman-diagrams the sum of which is proportional to the following
expression:
\bea
\rho &=& h^a_{\alpha\rho}\left((C^{1\alpha}T^b)_{ij}(C^{2\rho}T^b)_{kl}-
(C^{1\alpha}T^b)_{ij}(T^bC^{2\rho})_{kl}+ \right. \nn \\
&& \left. (T^bC^{1\alpha})_{ij}(T^bC^{2\rho})_{kl}-
(T^bC^{1\alpha})_{ij}(C^{2\rho}T^b)_{kl}\right),
\label{graphs}
\eea 
where $i..l$ are color indices.  If one treats $C^1$ and $C^2$ as
tensor operators acting 
on $R$ and recalls their commutation relation with $T^b$ this can be
rewritten in terms of the generators $\tau^{1b}$ and $\tau^{2b}$ on
representations $R_1$ and $R_2$, respectively:
\bea
\rho &=& h^a_{\alpha\rho}(\tau^{1b})^\alpha_{\alpha'}(\tau^{2b})^\rho_{\rho'}
       (C^{1\alpha'})_{ij}(C^{2\rho'})_{kl} \nn \\
     &=&-\half(C_{R_1}+C_{R_2}-C_{adj})h^a_{\alpha\rho}
(C^{1\alpha})_{ij}(C^{2\rho})_{kl},
\eea
where, we also used the tensor transformation properties of $h^a$.
Thus the operator renormalizes multiplicatively at one-loop order and its
anomalous dimension 
is determined by the quadratic Casimirs $C_{R_1}$, $C_{R_2}$, $C_{adj}$, of
representations $R_1$, $R_2$ and the adjoint:
\bea
\eta_1 = \frac{1}{2\pi}(C_{R_1}+C_{R_2}-C_{adj}),
\eea
and this together with the one-loop $\beta$-function, $\beta_0=
-\frac{C_{adj}}{4\pi}$ yields our magic number, $\alpha$:
\bea
\alpha = \frac{C_{R_1}+C_{R_2}-C_{adj}}{C_{adj}}
\label{alpha}
\eea
Equation \myref{alpha} is the main technical result of this paper. The power 
of this simple formula resides in that, $\alpha$ - which determines the order at
which the needed perturbative calculation becomes ``exact'' - may be such a
small integer that this calculation can be done in finite amount of time (as
in \cite{ope1}, where it was 1) or hopefully non-integer, in which case simply
no real computation is needed. Notice furthermore that, since \myref{graphs}
contains the flavor indices in a trivial way, the multiplicative
renormalization is 
also true for operators with nontrivial flavor-structure (in the multiflavor
models) and the same formula applies to their anomalous dimension.  

Let us summarize the above in the following recipe. Take a CGN-model, which
is defined by {\cal G}, $R$ and the number of flavors. Decompose the
product of representations $(\bar{R}\otimes R)\otimes(\bar{R}\otimes R)$ into
irreducibles 
and find the adjoints; following the proposed decomposition, to every copy of
the adjoint representation correspond two other ones, $R_1$ and $R_2$. Calculate
$\alpha_i$ using \myref{alpha} for every case obtained and take the highest
nonnegative integer, $\alpha$ out of them. Calculate the OPE
\myref{general} up to 
$\alpha$ loops perturbatively and see whether it closes on the color currents
or not; the fact you obtain is exact.


{\bf Examples.} We now consider applications of \myref{alpha} to various CGN
models and look for the ones where no real computation is needed. 
In ref. \cite{ope1} we calculated the OPE-coefficients up to $g^2$ in
perturbation theory. Though we kept the $SU(n)$-models in mind, the
computation is identical for any group and representation, and the statement
that {\em the OPE closes on the currents themselves up to 1-loop order} is
valid in all (M)CGN models. Therefore in the models under consideration, the 
conservation of the non-local charge is proved whenever among the $\alpha_i$'s
there is no integer greater than one. 

As a warm-up we repeat the result in the $SU(n)$ theories with the
fermions being in the fundamental representation. The
decomposition of the direct product is:
\bea
(\bar{R}\otimes R)\otimes(\bar{R}\otimes R) = 
(\unit\otimes R_{adj})\oplus(R_{adj}\otimes\unit)\oplus(R_{adj}\otimes R_{adj})
\oplus(\unit\otimes\unit).
\eea
The corresponding Casimirs are $C_\unit = 0 ; C_{adj} = 1$, which yields 
$\alpha_{\unit,adj} = 0$ and $\alpha_{adj,adj} = 1$.
The largest integer is 1, this was why we performed the one-loop
computation in \cite{ope1} and found that the quantum charge is conserved
in the multiflavor $SU(n)$-models. (In \cite{ope1} other
arguments, like C-parity were also used to rule out quadrilinear operators,
which is not needed here since their $\alpha$ are smaller than the largest
allowed one.)

Now consider the $SO(n)$ models with the fermions being in the vector
representation. As we expect, here we face more operators than in the
$SU(n)$-case. We repeat the decomposition for $SO(n)$:
\bea
(\bar{R}\otimes R)\otimes(\bar{R}\otimes R) &=&
(\unit\otimes R_{adj})\oplus(R_{adj}\otimes\unit)\oplus(R_{adj}\otimes R_{adj})
\oplus\nn \\
&&\oplus(R_S\otimes R_{adj})\oplus(R_{adj}\otimes R_S)\oplus(R_S\otimes R_S)
\oplus\ldots 
\eea
In this case $R=\bar{R}$ and $R_S$ stands for the symmetric tensor
representation, and we did not list the representations not containing
the adjoint here. Furthermore the Casimirs, $C_\unit = 0; C_{adj} = n-2; C_S =
n$ give the following $\alpha$'s:  
\bea
\alpha_{\unit,adj} = 0, \;\;\; \alpha_{adj,adj} = 1, \;\;\;
\alpha_{S,adj} = \frac{n}{n-2}, \;\;\; \alpha_{S,S} = \frac{n+2}{n-2}.
\eea
The largest integer is 5, 3 and 2 for $n=$3, 4 and 6, respectively and 1 in
all other cases, which proves the vanishing of the OPE-coefficient of extra
operators for almost every $n$; moreover under the mildest assumption about the
continuity of the crucial coefficient in terms of $n$ one conjectures
its vanishing 
for all $n$. This proves the conjecture that the $SO(n)$ CGN-models
also possess 
the Yangian algebra \cite{Bernard} (which can be used to prove their
integrability) and this is equally true for the multiflavor models as well. 
We can also consider other representations: as an example take the exotic
$SO(7)$ model with the fields in the $8$ spinor representation. One finds that
all the $\alpha$'s are non-integers except a 1, thus this model also possesses
Yangian symmetry. 

Let us now see how the procedure works for other groups taking $F_4$
first as an example. Let the fermions be in representation $26$ which is
self-conjugate and 
the bilinears decompose according to $26\otimes 26 = 1\oplus 26\oplus
52\oplus 273\oplus 324$. The quadratic Casimirs are $0,12,18,24,26$,
respectively and the corresponding
$\alpha$'s are summarized in the following table:

\begin{tabular}{|c||c|c|c|c|c|}\hline
$C_1\otimes C_2$ & $1\otimes 52$ & $26\otimes 26$ & $26 \otimes 273$ & $52
\otimes 52$ & $52\otimes 324$ \\ \hline
$\alpha$ & $0$ &$\frac{1}{3}$& $1$ &$ 1$ & $\frac{13}{9}$ \\ \hline 
\end{tabular}

\begin{tabular}{|c||c|c|c|}\hline
$C_1\otimes C_2$  
& $ 273\otimes 273$ & $273\otimes 324$ & $324\otimes 324$ \\ \hline
$\alpha$ & $\frac{5}{3}$ & $\frac{16}{9}$ & $\frac{17}{9}$ \\ \hline 
\end{tabular}

We can see that the biggest integer is 1 from which one concludes the absence
of extra operators in the OPE and the conservation of the non-local charge in
the $F_4$-model. It is straightforward to repeat the argument for other groups
and show that the same conclusion can be drawn for the {\cal G}$=E_7,
R=56$-model however in case of  {\cal G}$=E_6,R=27$ and {\cal G}$=E_8,R=248$
we obtain $\alpha_{650\otimes 650}=2$ and $\alpha_{30380\otimes 30380}=3$,
respectively, that is, a two and three-loop perturbative
calculation is needed in these models. It is clear that one can go on and play
around with 
other groups and representations using a table of group dimensions and indices
without difficulty. Note however that, to prove the {\em non-conservation} of
the non-local charge in a specific model one can not avoid going beyond
one-loop order in perturbation theory. 


{\bf Conclusion.} Once a theory possesses ``non-local'' charges
besides the usual ``local'' ones, it 
can be shown that, a Yangian symmetry algebra is generated. Due
to the non-local nature of the generators however, this is not a
direct multiplier of the 
two-dimensional Poincare group \cite{Lerr,Bernard} . This fact reduces the
dynamical question about the mass spectrum of a 
field theory to one on the classification of representations of the underlying
algebra (see e.g. \cite{Belavin,Nakanishi}); and after identifying 
the spectrum, the S-matrix is also highly constrained being proportional to
the R-matrix in the given representation. Another
approach to the determination of the S-matrix, (which was originally
developed by L\"uscher) uses the fact that the action of the non-local
charge on asymptotic one-particle  
states is straightforwardly computed, in terms of which all the
asymptotic matrix elements can be obtained \cite{Lerr,Buchholz}. These
results lead to the absence of particle production and factorization
\cite{Luscher,Abdalla}.

In this letter, various chiral Gross-Neveu models -- which are identified by
their symmetry group, representation and multiplicity of the fields -- have
been considered, and the existence of the non-local generator of the expected
Yangian symmetry algebra was discussed. It has long been known that the
$SU(n)$ one-flavor model possess the Yangian symmetry but one expects that
there may be other theories with this property, too. In a previous article we
proved this for the multiflavor $SU(n)$ model and the method developed there was
extended to the whole family in the present paper. The intricate question
related to the non-perturbative definition of the model could be answered using
finite order perturbation theory (thanks to asymptotic freedom) -- often a
one-loop calculation is sufficient. Examples for the latter case were
presented proving among others that in the $SO(n)$ models the
non-local charge can also be defined and that, more exotic theories possess
it as well; moreover that, this property is independent of the number of
flavors. We also have found CGN-models where -- within this framework
-- only a higher-loop calculation
could decide which class they belong to; and it is an open question
whether there exist generalized CGN quantum theories without the
non-local charge generating Yangian symmetry at all.  


{\bf Acknowledgment.} I am grateful to J. Balog and P. Forg\'acs for helpful
discussions. This work was partly supported by the Hungarian National Science
Fund OTKA, grant No. T19917.


\end{document}